\documentclass[a4paper,11pt,english]{article}
\usepackage[utf8]{inputenc}
\usepackage{babel,amsmath,amssymb,amsthm,enumitem,hyperref,cleveref}
\usepackage[sort&compress,numbers,square,sort]{natbib}

\renewcommand{\P}{\mathrm{P}}
\newcommand{\Q}{\mathrm{Q}}
\newcommand{\E}{\mathrm{E}}
\newcommand{\e}{\mathcal{E}}
\newcommand{\I}{\mathrm{I}}
\newcommand{\B}{\mathcal{B}}
\newcommand{\F}{\mathcal{F}}
\newcommand{\PP}{\mathcal{P}}
\newcommand{\M}{\mathcal{M}}
\newcommand{\PPt}{\mathcal{\tilde P}}
\newcommand{\FF}{\mathbb{F}}
\newcommand{\R}{\mathbb{R}}
\newcommand{\cadlag}{{c\`adl\`ag}}

\renewcommand{\hat}{\widehat}
\renewcommand{\tilde}{\widetilde}
\newcommand{\midd}{\;\bigg|\;}
\let\Ss\S
\renewcommand{\S}{\mathcal{S}}
\renewcommand{\epsilon}{\varepsilon}
\newcommand{\cint}{\boldsymbol{\cdot}}
\DeclareMathOperator{\lln}{\overline{ln}}

\newtheorem{theorem}{Theorem}
\newtheorem{lemma}{Lemma}
\newtheorem{proposition}{Proposition}

\theoremstyle{definition}
\newtheorem{remark}{Remark}

\title{Survival investment strategies in a continuous-time market model with
competition}
\author{Mikhail Zhitlukhin\thanks{Steklov Mathematical Institute of the
Russian Academy of Sciences.
8 Gubkina St., Moscow, Russia. Email: mikhailzh@mi-ras.ru.
The research was supported by the Russian Science Foundation,
project no. 18-71-10097.
}}
\date{4 September 2019}

\begin{document}
\maketitle

\begin{abstract}
We consider a stochastic game-theoretic model of an investment market in
continuous time with short-lived assets and study strategies, called
survival, which guarantee that the relative wealth of an investor who uses
such a strategy remains bounded away from zero. The main results consist in
obtaining a sufficient condition for a strategy to be survival and showing
that all survival strategies are asymptotically close to each other. It is
also proved that a survival strategy allows an investor to accumulate wealth
in a certain sense faster than competitors.

\end{abstract}

\section{Introduction}

This paper proposes a stochastic game-theoretic model of an investment
market in continuous time where investors compete for payoffs yielded by
several assets. The main objective is to study questions about asymptotic
optimality of investment strategies from \emph{evolutionary} point of view:
to describe what strategies \emph{survive} in the competition for payoffs,
what strategies \emph{dominate} or \emph{get extinct}, and how they affect
the market structure in the long run. This circle of questions has been
studied in a number of papers in the literature, but mostly in discrete time
(see the reviews~\cite{EvstigneevHens+16,Holtfort19}, and
\Cref{section-review} below). The model considered here is one of the few in
continuous time.

A market in our model consists of several investors who invest their wealth
in assets. Asset payoffs, which are specified by some exogenous stochastic
processes and paid continuously, are distributed between the investors
proportionally to the amount of wealth they allocate to each asset.
Naturally, a larger expected future payoff of an asset will attract more
investors, which will reduce the share of the payoff received by each of
them. Hence the investors face the problem how to allocate their wealth in
an optimal way. In our model, assets are assumed to be short-lived in the
sense that they are bought by investors, yield payoffs at the ``next
infinitesimal'' moment of time, and then reappear again, but cannot be sold
to capitalize on increased prices (such a model is simpler than a model of a
stock market).

One of the main goals of the paper is to identify strategies that
survive in the market in the sense that the relative wealth of an
investor who uses such a strategy remains bounded away from zero on the
whole time axis (by the relative wealth we mean the share of wealth of one
investor in the total wealth of the market). It is not assumed that all the
investors are rational, i.e.\ that their strategies are defined as solutions
of some optimization problems. For example, they can use strategies that
mimic other market participants, follow some empirical rules, etc. It is
also not assumed that the investors know the strategies of their
competitors. Therefore, a survival strategy should be robust in the sense
that it guarantees a positive share of wealth no matter what strategies are
used by the other investors.

The main results of the paper are as follows. First, we obtain a sufficient
condition for a strategy to be survival. It is stated in an explicit form:
we construct one particular survival strategy and show that any other
strategy which is asymptotically close to it is survival as well. We also
prove that such a strategy dominates in the market, i.e.\ the relative wealth
of an investor who uses it tends to 1, when the representative strategy of
the other investors is asymptotically different from it. Moreover, we show
that using a survival strategy allows to achieve the highest asymptotic
growth rate of wealth among the investors in the market.

The key idea to obtain these results is to find a strategy such that the
process of the logarithm of its relative wealth is a submartingale. As it
will be shown, its existence follows from Gibbs' inequality applied to a
suitable representation of the relative wealth process. The survival
property is then established using results on convergence of submartingales.

This approach was used for the first time in the paper
\cite{AmirEvstigneev+13}, which studied a fairly general discrete-time model
with short-lived assets. For particular instances of that model, similar
results had been known before (see the review \cite{EvstigneevHens+16}), but
they mainly used ideas based on the Law of Large Numbers, which limited them
only to payoff sequences consisting of independent random variables. One can
also mention the paper \cite{AmirEvstigneev+11}, where an approach similar
to \cite{AmirEvstigneev+13} was used in a model with long-lived assets,
which describes a usual stock market. There, in order to obtain similar
results, more subtle arguments were required and some restrictive
assumptions were imposed on the model. From this point of view, our work is
closer to \cite{AmirEvstigneev+13}, we also consider a model with only
short-lived assets, but in continuous time.

Let us also mention that survival strategies are somewhat similar to growth
optimal strategies in asset market models without competition (see
\cite{HakanssonZiemba95,PlatenHeath06,KaratzasKardaras07}), as both arise
from the problem of maximizing the logarithm of wealth. In particular, we
show that survival strategies allow to achieve the highest growth rate of
wealth, similarly to growth optimal strategies. However, an essential
difference between these two classes of strategies is that survival
strategies cannot be directly obtained from a single-agent wealth
optimization problem because the evolution of wealth of one investor depends
also on actions of the other investors.

The paper is organized as follows. In \Cref{section-model}, we describe the
model. In \Cref{section-survival}, we introduce the notion of a survival
strategy and provide an explicit construction of one such strategy. The
main results of the paper are stated in \Cref{section-main-results}.
\Cref{section-proofs} contains their proofs.

\section{The market model}
\label{section-model}

Before we study a general model in continuous time, let us consider a model
in discrete time, in which the main objects and formulas have a clear
interpretation. Based on it, we will formulate the general model.

\subsection{Preliminary consideration: a model in discrete time}
\label{section-discrete-model}

Let us fix a probability space $(\Omega,\F,\FF,\P)$ with a discrete-time
filtration $\FF = (\F_t)_{t=0}^\infty$, on which all the random variables
will be defined.

The market in the model includes $M\ge 2$ investors and $N\ge2$ assets,
which yield non-negative random payoffs at moments of time $t=1,2,\ldots$
The investors decide, simultaneously and independently of each other at
every moment of time, what part of their wealth they invest in each of the
assets, and the asset payoffs are split proportionally to the invested
wealth amounts. The investment decision are made before the payoffs become
known. We impose the assumption that at every moment of time the proportions
of own wealth the investors allocate to the assets are the same for all the
investors (however the distribution of invested wealth between the assets
may be different and the investors are free to choose it); a model where
they may differ would be more complicated and is not studied in this paper.

The payoffs of the assets are specified by random sequences $A_t^n(\omega)
\ge 0$, which are adapted to the filtration ($A_t^n$ is $\F_t$-measurable).
These random sequences are exogenous, i.e.\ do not depend on actions of the
investors.

The evolution of the investors' wealth is described by adapted sequences
$Y_t^m(\omega)\ge 0$. The initial values $Y_0^m>0$ are given, and further
values depend on strategies used by the investors. A strategy of investor
$m$ is identified with a sequence $\lambda_t^m(\omega)\ge 0$ of random
vectors in $\R^N$, which express the proportions of wealth invested in each
of the assets. The sequences $\lambda^{m,n}$ are predictable
($\lambda_t^{m,n}$ if $\F_{t-1}$-measurable) and $\sum_n\lambda_t^{m,n} =
1$.

Given this, we state the equation which determines the evolution of investor
$m$'s wealth:
\begin{equation}
Y_t^m = (1-\delta_t) Y_{t-1}^m + \sum_n \frac{\lambda_t^{m,n}Y_{t-1}^{m}}{\sum_k
\lambda_t^{k,n} Y_{t-1}^k} A_t^n, \qquad t\ge 1,\label{wealth-discrete-1}
\end{equation}
where $\delta_t(\omega)$ is the proportion of wealth each investor allocates
for investment in the assets. The sequence of random variables
$\delta_t\in[0,1)$ is predictable, given exogenously, and the same for all
the investors.

Notice that the first term in the right-hand side
of~\eqref{wealth-discrete-1} is the amount of wealth not invested in the
assets, and the second term is the received payoff. The fraction in the sum
expresses the idea of division of payoffs proportionally to invested amounts
of wealth. We treat the indeterminacy 0/0, which happens when no one invests
in asset $n$, as $0/0=1/M$, so in this case the payoff of the corresponding
asset is split in the equal proportions. Note that always $Y_t^m>0$ due to
the assumption $\delta_t<1$.

Let us emphasize that the components of the strategies $\lambda_t^{m,n}$
depend on a random outcome $\omega\in\Omega$, but do not depend on the
investors' wealth or their strategies. This means that the investors, when
deciding how to allocate their wealth, take into consideration only asset
payoffs. Such strategies can be called \emph{basic} (as, e.g., in the paper
\cite{AmirEvstigneev+13}). One could consider a more general model, where,
for example, $\lambda_t^m = \lambda_t^m(\omega,Y_0,\ldots,Y_{t-1},
\lambda_0,\ldots,\lambda_{t-1})$, but this will not essentially increase the
generality of the main results of our paper, see
Remark~\ref{remark-general-strategies} below.

In order to get an idea how to state a continuous-time
counterpart of equation~\eqref{wealth-discrete-1}, let us rewrite
it in the following form:
\begin{equation}
\label{wealth-discrete-2}
\Delta Y_t^m = - Y_{t-1}^m \Delta V_t + \sum_n \frac{\lambda_t^{m,n}Y_{t-1}^{m}}{\sum_k
\lambda_t^{k,n} Y_{t-1}^k} \Delta X_t^n, \qquad t\ge 1,
\end{equation}
where $X_t,V_t$ are the sequences of cumulative payoffs and cumulative
investment proportions defined as
\begin{equation}
X_t = \sum_{s\le t}A_s,\qquad V_t = \sum_{s\le t} \delta_s ,\label{V-X}
\end{equation}
and the symbol $\Delta$ denotes a one-period increment, e.g.\ $\Delta Y_t^m =
Y_t^m - Y_{t-1}^m$.

The form of equation~\eqref{wealth-discrete-2} suggests that an analogous
model in continuous time can be obtained by considering continuous-time
processes $X_t$, $V_t$, $Y_t$ and ``replacing'' the one-step increments with
the infinitesimal increments, e.g.\ $\Delta Y_t$ with $dY_t$. Our goal for
the rest of this section will be to define such a model in a proper way.

\subsection{Notation}

Let us introduce notation that will be used to formulate the continuous-time
version of the above model.

From now on assume given a filtered probability space $(\Omega,\F,\FF,\P)$
with a continuous-time filtration $\FF=(\F_t)_{t\in\R_+}$, which satisfies
the usual assumptions, i.e.\ $\FF$ is right-continuous ($\F_t =
\bigcap_{s>t} \F_s$), the $\sigma$-algebra $\F$ is $\P$-complete, and $\F_0$
contains all the $\P$-null sets of $\F$.

For vectors $x,y\in \R^N$, by $xy$ we will denote the scalar product, by
$|x|$ the $l_1$-norm of a vector, and by $\|x\|$ the $l_2$-norm; for a
scalar function $f\colon \R\to\R$ the notation $f(x)$ means the
coordinatewise application of the function:
\[
\begin{split}
&xy = \sum_n x^ny^n,\quad |x|= \sum_n |x^n|, \quad  \|x\| = \sqrt{xx}, \\
&f(x) = (f(x^1),\ldots, f(x^N)).
\end{split}
\]
If $G_t=G(t)$ is a non-decreasing function, then for a measurable function
$f_t$ denote
\[
f\cint G_t = \int_0^t f_s d G_s,
\]
provided that the integral is well-defined (as a Lebesgue-Stieltjes
integral). Functions $f,G$ may be random, then $f\cint G_t(\omega)$ is
defined pathwise for each~$\omega$. If $f$ is vector-valued and
$G$ is scalar-valued, then $f\cint G_t = (f^1\cint G_t,\ldots,f^N\cint
G_t)$; if both are vector-valued, then $f\cint G_t = \sum_n f^n \cint
G^n_t$.

As usual, all equalities and inequalities between random variables are
assumed to hold with probability one (almost surely). For random processes
$X_t(\omega)$, $Y_t(\omega)$, the equality $X=Y$ is understood to hold up to
$\P$-indistinguishability, i.e.\ $\P(\exists\, t : X_t\neq Y_t) = 0$; in the
same way we treat inequalities. Properties of trajectories (continuity,
monotonicity, etc.) are assumed to hold for all $\omega$, unless else is
specified. If $X,Y$ are right-continuous processes, then $X=Y$ if and only
if $X_t=Y_t$ a.s.\ for all $t$.

By the predictable $\sigma$-algebra $\PP$ on $\Omega\times\R_+$ we call, as
usual, the $\sigma$-algebra generated by all left-continuous adapted
processes. A process is predictable if it is measurable with respect to
$\PP$ as a map from $\Omega\times\R_+$ to $\R$ or to $\R\cup\{\pm\infty\}$.

\subsection{The general model}

As in the discrete-time model, there are $M\ge 2$ investors and $N\ge 2$
assets. The asset payoffs are specified by exogenous cumulative payoff processes
$X_t^n$ (cf.~\eqref{V-X}), which are adapted to the filtration $\FF$ and
have non-decreasing \cadlag\ paths (right-continuous with left limits) with
$X_0^n=0$.

The cumulative proportion of wealth allocated by each investor to the assets
is specified by means of an adapted non-decreasing \cadlag\ scalar process
$V_t$ with $V_0=0$ and jumps $\Delta V_t \in[0,1)$ (as usual, $\Delta V_t =
V_t - V_{t-}$, where $V_{t-} = \lim_{s\uparrow t} V_s$, and $\Delta V_0=0$).
To avoid problems with non-integrability (see
\Cref{section-survival-construction}), we will assume that the jumps of $V$
are uniformly bounded away from 1, i.e.\ there exists a constant $\gamma_V
\in [0,1)$ such that for all $\omega\in\Omega$ and $t\ge 0$
\begin{equation}
\Delta V_t(\omega) \le \gamma_V.
\label{V-jumps-bound}
\end{equation}

A strategy of investor $m$ is identified with a predictable process
$\lambda_t^m$ of proportions of wealth invested in the assets, which assumes
values in the standard simplex in $\R^N$, i.e.\ $\lambda_t^{m,n}\ge 0$ and
$\sum_n \lambda_t^{m,n}=1$. As was noted above, we consider only basic
strategies, which means that $\lambda_t^m$ does not depend on the ``past
history'' of the market.

The wealth processes of the investors are defined as strictly positive
\cadlag\ processes $Y^m$ that satisfy the equation (a continuous-time
counterpart of \eqref{wealth-discrete-2})
\begin{equation}
d Y_t^m = -Y_{t-}^md V_t + \sum_n \frac{\lambda_t^{m,n}Y_{t-}^m}{\sum_k \lambda_t^{k,n}
Y_{t-}^k} d X_t^n
\label{wealth-sde}
\end{equation}
and such that $Y_->0$ (i.e.\ $Y_{t-}^m >0$ for all $t\ge0$ and $m$). Without
loss of generality, we will always assume that the initial values $Y_0^m>0$
are non-random. If $\lambda_t^{k,n} =0$ for all $k$, then we assume that the
value of the fraction in the right-hand side is equal to $1/M$ for
corresponding $n$.

As usual, equation~\eqref{wealth-sde} should be understood in the integral
form:
\begin{equation}
Y_t^m = Y_0^m +  \sum_n \int_0^t \frac{\lambda_s^{m,n}Y_{s-}^m}{\sum_k
\lambda_s^{k,n} Y_{s-}^k} d X_s^n - \int_0^t Y_{s-}^m d V_s, \qquad t\ge 0.
\label{wealth}
\end{equation}
The integrals here are pathwise Lebesgue-Stieltjes integrals (which are
well-defined since the processes $X_t$, $V_t$ do not decrease, and the
integrands are non-negative). It is not difficult to see that if $Y^m$
satisfies \eqref{wealth}, then it has finite variation on any interval
$[0,t]$.

The next proposition shows that equation~\eqref{wealth-sde} has a unique
solution, hence the wealth processes are well-defined.

\begin{proposition}
\label{proposition-correctness}
For any non-random initial capitals $Y_0^m>0$ and strategies $\lambda^{m}$,
$m=1,\ldots,M$, there exists a unique adapted strictly positive \cadlag\
process $Y=(Y^1,\ldots,Y^M)$ which satisfies \eqref{wealth} and $Y_- > 0$.
\end{proposition}

\section{Survival strategies}
\label{section-survival}

\subsection{The notion of survival}

For given initial capitals $Y_0^m$, investment strategies $\lambda^m$, and the
corresponding wealth processes $Y^m$, define the process of total
market wealth $W$ and the relative wealth $r^m$ of investor $m$:
\[
W_t = |Y_t|, \qquad r_t^m = \frac{Y_t^m}{W_t}.
\]
In the case when it is necessary to emphasize that the introduced processes
depend on the initial capitals and the strategies, we will use the notation
$Y_t^m(Y_0,\Lambda)$ and $r_t^m(Y_0,\Lambda)$, where $\Lambda =
(\lambda^1,\ldots,\lambda^M)$ denotes a profile of strategies.

The central definition of the present paper is the notion of a
\emph{survival strategy}. We call a strategy $\lambda$ survival, if for any
initial capitals $Y^m_0>0$, $m=1,\ldots,M$, and a strategy profile $\Lambda
= (\lambda^1,\ldots,\lambda^M)$ with $\lambda^1=\lambda$ and arbitrary
strategies $\lambda^m$, $m=2,\ldots,M$, with probability one it holds that
\[
\inf_{t\ge0} r_t^1(Y_0, \Lambda) > 0,
\]
i.e.\ a survival strategy guarantees that an investor who uses it will
always have a share in the total wealth bonded away from zero.

As observed above, $Y>0$ and $Y_->0$, hence the notion of survival can be
equivalently stated as that $\liminf_{t\to\infty} r_t^1(Y_0,\Lambda)>0$.

Equivalently, it can be also reformulated as the property that there exists
a strictly positive random variable $C$ (generally, depending on the initial
capitals $Y_0^m$ and the strategy profile $\Lambda$) such that
\[
Y_t^1 \ge CY_t^m\ \text{for all}\ m\ \text{and}\ t\ge0,
\]
i.e.\ no strategy can provide the asymptotic growth of wealth faster than a
survival strategy.

\subsection{Construction of a survival strategy}
\label{section-survival-construction}

We will now explicitly construct a candidate survival strategy. Its survival
property, as well as other asymptotic optimality properties, will be
established in \Cref{section-main-results}. The exposition below relies on
several known facts from stochastic calculus, which can be found, for
example, in \cite{JacodShiryaev02}.

Let us split the process $X_t$ into the continuous part $X_t^c$ and the sum
of jumps, i.e.
\[
X_t = X_{t}^{c} + \sum_{s\le t} \Delta X_s,
\]
where $X_t^c$ is a continuous non-decreasing process, $\Delta X_s = X_s -
X_{s-}$, and for $s=0$ we set $\Delta X_0 = 0$. It will be convenient to
work with jumps $\Delta X_t$ and $\Delta V_t$ using the measure of jumps of
the $(N+1)$-dimensional process $(X_t,V_t)$. It is defined as the
integer-valued random measure on $(\mathcal{S}, \B(\S))$, where $\S =
\R_+\times \R^{N+1}_+$ and $\B$ is the Borel $\sigma$-algebra, by the
formula
\[
\mu(\omega, A) = \sum_{t\ge 0} \I(\Delta (X_t, V_t)(\omega)\neq 0,\;
(t,\Delta (X_t, V_t)(\omega)) \in A), \qquad A\in \B(\S)
\]
(actually, we can assume $\S = \R_+\times\R^N_+\times [0,\gamma_V]$, where
$\gamma_V$ is the constant from bound \eqref{V-jumps-bound}). For an
integral of an $\F\otimes\B(\S)$-measurable function with respect to a
random measure we will use the notation
\begin{equation}
f*\mu_t(\omega) = \int_{(0,t]\times \R^{N+1}_+} f(\omega,s,x,v)
\mu(\omega,ds,dx,dv),\label{random-measure-integral}
\end{equation}
assuming that the integral is well-defined (as a Lebesgue integral),
possibly being $+\infty$ or $-\infty$. Henceforth, the variable $x\in
\R^N_+$ corresponds to jumps of $X$, and $v\in \R_+$ to jumps of $V$. The
integral \eqref{random-measure-integral} can be defined for a general random
measure; in the particular case when $\mu$ is the measure of jumps of $(X,V)$, it can
be simply written as the sum
\[
f*\mu_t(\omega) = \sum_{s\le t} f(\omega, s, \Delta X_s(\omega), \Delta V_s(\omega))
\I(\Delta (X_s,V_s)(\omega)\neq 0).
\]
In the case when $f$ is a vector-valued measurable function, we treat the
integral \eqref{random-measure-integral} as vector-valued and compute it
coordinatewisely. In particular, the process $X_t$ can be represented in the
form
\[
X_t = X_t^c + x * \mu_t,
\]
where $x=(x^1,\ldots,x^N)$ and $x^n$ stands for the function $(x,v)\mapsto
x^n$.

Let $\PPt = \PP\otimes \B(\R^{N+1}_+)$ be the predictable $\sigma$-algebra
on $\Omega\times\R_+\times\R^{N+1}_+$.
Recall that a random measure $\nu$ is called predictable, if for any
$\PPt$-measurable non-negative function $f(\omega,t,x,v)$ the process
$f*\nu_t$ is predictable ($\PP$-measurable). From now on, let $\nu$ denote
the compensator of the measure of jumps $\mu$, i.e.\ a predictable random
measure such that for any $\PPt$-measurable non-negative function $f$ holds
the equality
\[
\E (f* \mu_\infty ) = \E(f*\nu_\infty),
\]
or, equivalently, $f*(\mu-\nu)_t$ is a local martingale, provided that the
process $|f|*\mu_t$ is locally integrable.
The measure of jumps of an adapted \cadlag\ process always has a
compensator, which is unique up to indistinguishability with respect to $\P$
\cite[\Ss\,II.1]{JacodShiryaev02}, hence in our model $\nu$ is
well-defined.
Since the processes $X$ and $V$ do not decrease, the inequality $(|x|\wedge
1+v)*\nu_t <\infty$ holds a.s.\ for all $t$, see
\cite[\Ss\,4.1]{LiptserShiryaev89en}.

From the general theory, it is known that there exists a predictable
\cadlag\ non-decreasing locally integrable scalar process $G$ (an
\emph{operational time process}) such that, up $\P$-indistinguishability,
\begin{equation}
X_t^c = b\cint G_t, \qquad \nu(\omega,dt,dx,dv) = K_{\omega,t}(dx,dv)
dG_t(\omega),\label{operation-time}
\end{equation}
where $b_t$ is a predictable process with values in $\R_+^N$, and 
$K_{\omega,t}(dx,dv)$ is a transition kernel from $(\Omega\times\R_+,\PP)$ to
$(\R^{N+1}_+,\B(\R^{N+1}_+))$ which for all $\omega,t$ satisfies the properties
\[
K_{\omega,t}(\{0\})=0, \qquad
\int_{\R^{N+1}_+} (|x|\wedge 1+v) K_{\omega,t}(dx,dv) < \infty.
\]
Fore example, one can use the process
\begin{equation}
G_t = |X_t^c| + (|x|\wedge 1+v) * \nu_t\label{G-canonical}.
\end{equation}
The possibility of representation \eqref{operation-time} for this process
can be proved similarly to Proposition~II.2.9 in \cite{JacodShiryaev02}.

For $b,K,G$ satisfying \eqref{operation-time}, define the predictable
process $a_t$ with values in $\R^N_+$ by the formula
\begin{equation}
a_t^n(\omega) = \int_{\R^{N+1}_+} \frac{x^n}{1-v +|x|/W_{t-}(\omega)}
K_{\omega,t}(dx,dv),\label{a-def}
\end{equation}
and define the  strategy $\hat\lambda$ by
\begin{equation}
\hat\lambda_t = \frac{a_t + b_t}{|a_t+b_t|},\label{survival}
\end{equation}
where we put $\hat\lambda_t^n=1/N$ for all $n$ whenever $|a_t+b_t|=0$. This
strategy will be a candidate for a survival strategy. Note that the
continuous part of the process $V$ is not involved in its construction.

Observe that the strategy $\hat\lambda$ does not essentially depend on the
choice of an operational time process in the following sense. Let the
process $G$ be defined by \eqref{G-canonical}. Define the measure
$\Q=\P\otimes G$ on $(\Omega\times \R_+, \PP)$, i.e.\ for $A\in\PP$
\[\Q(A) = \E\biggl(\int_{0}^\infty \I((\omega,t)\in A)
dG_t(\omega)\biggr).
\]
\begin{proposition}
\label{prop-equal-realizations}
Suppose $G'$ is another operational time process satisfying
\eqref{operation-time} and let $\hat\lambda,\hat\lambda'$ be the strategies
constructed with respect to $G,G'$ as described above. Then $\hat\lambda =
\hat\lambda'$ ($Q$-a.s.).

Moreover, for any initial capitals $Y_0^m>0$, $m=1,\ldots,M$, and strategy
profiles $\Lambda=(\hat\lambda,\lambda^2,\ldots,\lambda^M)$, $\Lambda' =
(\hat\lambda',\lambda^2,\ldots,\lambda^M)$ with arbitrary strategies
$\lambda^m$, $m=2,\ldots,M$, the corresponding wealth processes are equal,
i.e.\ $Y(Y_0,\Lambda) = Y(Y_0,\Lambda')$.
\end{proposition}

\begin{remark}
Obviously, the discrete-time model of \Cref{section-discrete-model} is a
particular case of the general model. In discrete time, $\Delta X_t = A_t$,
$\Delta V_t = \delta_t$, and one can take $G_t=[t]$ (the integer part). Then
$X_t^c=0$, and $K_{\omega,t}(dx,dv)$ is the regular conditional distribution
of the pair $(A_t, \delta_t)$ with respect to $\F_{t-1}$. By straightforward
computation, we find
\begin{equation}
a_t^n = W_{t-1} \E\biggl(\frac{A_t^n}{W_t} \;\biggl|\; \F_{t-1}\biggr), \qquad
b_t= 0, \qquad \hat\lambda_t = \frac{a_t}{|a_t|}.
\label{dicrete-survival}
\end{equation}
\end{remark}

\section{Main results}
\label{section-main-results}

\subsection{Statements}

In this section we assume given and fixed an operational time process $G$
for which representation \eqref{operation-time} holds, and $a,b,K, \hat
\lambda$ constructed from $G$ as described in the previous section. To
formulate the results, let us also introduce the predictable scalar process
\[
H_t = \frac{|a+b|}{W_-}\cint G_t.
\]
\begin{proposition}
\label{H-bound}
The process $H$ is finite: $H_t<\infty$ a.s.\ for all $t\ge 0$.
\end{proposition}

For an adapted scalar process $L_t$ we will use the notation
$\M(L)=\{\tau_l(L),\;l\in\R_+\}$ for the class of stopping times when $L$
exceeds a level~$l$ for the first time: $ \tau_l(L) = \inf\{t\ge 0: L_t \ge
l\}$, where $\inf\emptyset = +\infty$.

Our first main result formulated in the following \namecref{theorem-survival}
states that if a strategy $\lambda$ is close to $\hat\lambda$ in a certain
sense, then it is survival. In particular, $\hat\lambda$ itself is survival.

\begin{theorem}
\label{theorem-survival}
Suppose a strategy $\lambda$ satisfies the following conditions:
\begin{enumerate}[label=(\alph*),leftmargin=*,itemsep=0mm,topsep=1mm]
\item $\P(\exists\,t : \lambda_t^n =
0,\;\hat\lambda_t^n \neq 0)=0$ for all $n$,
\item the process $U_t = \hat\lambda_t(\ln \hat\lambda_t -
\ln\lambda_t)$ satisfies $U\cint H_\infty < \infty$,
\item $\E(U_\tau \Delta H_\tau \I(\tau<\infty))  <\infty$ for any
$\tau\in\M(U\cint H)$.
\end{enumerate}
Then, if investor $m$ uses the strategy $\lambda$, the limit
$\lim_{t\to\infty} r_t^m > 0$ exists with probability one for any strategies
$\lambda^k$ of the other investors. In particular, the strategy $\lambda$ is
survival.
\end{theorem}
The proximity of a strategy $\lambda$ to $\hat\lambda$ is essentially
determined by condition~(b), while (a) and (c) are technical assumptions.
Let us clarify that in conditions (b), (c) on the sets $\{\lambda_t^n=0\}$
and $\{\hat\lambda_t^n=0\}$ the corresponding term in the scalar product in
the definition of $U_t$ is assumed to be zero. Also observe that the process
$U$ is non-negative as follows from Gibbs' inequality (see also
\Cref{lemma-logsum} below). Therefore, the integral $U\cint H_\infty$ and
the expectation in condition (c) are always well-defined, though they may
take on the value $+\infty$.

The next simple proposition can be useful for verification of conditions
(a), (b) of \Cref{theorem-survival} in particular models. Regarding (c), a simple
sufficient condition for its validity is the continuity of the process $G$
(and, hence, the continuity of $H$). For example, $G$ is continuous when $X$
and $V$ are non-decreasing L\'evy processes.

\begin{proposition}
\label{prop-survival-sufficient}
Suppose the processes $X,V$ are such that the strategy $\hat\lambda$
satisfies the inequality $\inf_{t\ge0} \hat\lambda_t^n > 0$ for all $n$. In
that case, if a strategy $\lambda$ satisfies the inequalities $\inf_{t\ge 0}
\lambda_t^n > 0$ for all $n$ and $\|\hat\lambda- \lambda\|^2\cint H_\infty <
\infty$, then it satisfies conditions (a), (b) of \Cref{theorem-survival}.
\end{proposition}

The next result shows that all survival strategies are, in a certain sense,
asymptotically close to the strategy $\hat \lambda$.
\begin{theorem}
\label{theorem-close}
If a strategy $\lambda$ is survival, then $\|\hat
\lambda - \lambda \|^2 \cint H_\infty < \infty$.
\end{theorem}

The third theorem formulated below shows that a survival strategy dominates
in the market, i.e.\ the relative wealth of an investor who uses it tends to
1 as $t\to\infty$, if the \emph{representative strategy} of the other
investors is essentially different from $\hat \lambda$ in a certain sense.
By the representative strategy of investors $k\neq m$ we call the
predictable process $\tilde \lambda$ which is the weighted sum of the
strategies of these investors with their relative wealths as the weights:
\[
\tilde \lambda^{n}_t = \frac1{1-r_{t-}^m} {\sum_{k\neq m} \lambda_t^{k,n}
r_{t-}^k}.
\]
Notice that $|\tilde\lambda_t| = 1$.

\begin{theorem}
\label{theorem-dominate}
Suppose investor $m$ uses a strategy $\lambda$ which satisfies the
conditions of \Cref{theorem-survival}. Let $\tilde\lambda$ be the
representative strategy of the other investors. Then $\lim_{t\to\infty}r_t^m
= 1$ a.s.\ on the set $\{\|\hat\lambda - \tilde \lambda\|^2\cint H_\infty =
\infty\}$.
\end{theorem}

The last result draws parallels between survival strategies in our model and
growth optimal (or log-optimal) strategies in asset market models with
exogenous asset prices (see the discussion in the next section). To state
it, let us define the asymptotic growth rate of investor $m$'s wealth
$Y_t^m$ as
\[
\limsup_{t\to\infty} \frac1t {\ln Y_t^m}
\]
(similarly to the definition of the asymptotic growth rate in asset market
models with exogenous prices, see, e.g.,
\cite[Chapter~3.10]{KaratzasShreve98}), and define the growth rate of wealth
$Y_t^m$ between moments of time $s<t$ as
\[
\frac1{t-s}\E\biggl( \ln\frac{Y_t^m}{Y_s^m}\midd \F_s\biggr).
\]

\begin{theorem}
\label{theorem-opt-growth}
1) If investor $m$ uses a survival strategy, then this investor achieves the maximal
asymptotic growth rate of wealth in the market: for any $k$
\[
\limsup_{t\to\infty} \frac1t {\ln Y_t^m} \ge \limsup_{t\to\infty} \frac1t
{\ln Y_t^k}.
\]

\noindent
2) Suppose investor $m$ uses the strategy $\hat\lambda$ and let $\tilde Y_t
= \sum_{k\neq m} Y_t^k$ denote the total wealth of the other investors. Then
$Y_t^m$ grows faster than $\tilde Y_t$ between any moments of time $s< t$
such that $\E (|\ln W_t|\mid\F_s)<\infty$, i.e.
\[
\E\biggl( \ln\frac{Y_t^m}{Y_s^m}\midd \F_s\biggr) \ge \E\biggl(
\ln\frac{\tilde Y_t}{\tilde Y_s}\midd \F_s\biggr).
\]
\end{theorem}

Note that in the second claim of \Cref{theorem-opt-growth}, it is generally
not possible to say that $\E( \ln(Y_t^m/Y_s^m)\mid \F_s) \ge \E(
\ln(Y_t^k/Y_s^k)\mid \F_s)$ for any $k$ if the number of investors $M\ge 3$.

\begin{remark}
\label{remark-general-strategies}
As was noted above, all investment strategies considered in the present
paper are basic in the sense that their components are functions of $t$ and
$\omega$ only. One could also extend the model by allowing general
strategies, where $\lambda_t$ may depend on paths of the processes $Y$,
$\lambda$ up to time $t$ in an appropriate non-anticipative way. However,
most of the above results will remain valid in such an extended setting as
well. Let us give an heuristic argument for that without entering into
technical details.

For example, assume that the strategies $\lambda_t^m(\omega,Y_{t-}(\omega))$
can also depend on the current wealth in a way such that the wealth equation
admits a unique solution $Y$. Then we can consider the realizations of the
strategies $\bar \lambda_t^m(\omega) = \lambda_t^m(\omega,Y_{t-}(\omega))$
(provided that they are predictable processes), and by inspecting the
proofs, one can see that \Cref{theorem-survival,theorem-opt-growth,theorem-dominate} and
\Cref{prop-survival-sufficient} will remain valid, if it is additionally required that
a strategy $\lambda$ in their statements is basic. In particular,
\Cref{theorem-survival} implies that in a model with general strategies a survival
strategy exists and can be found among basic strategies ($\hat\lambda$ is
such a strategy). However, only basic survival strategies will be
asymptotically close to $\hat\lambda$, i.e.\ \Cref{theorem-close} does not hold
if one allows $\lambda$ to be a general survival strategy. A counterexample
is provided in the paper \cite{AmirEvstigneev+13} for a different model, but
it can be carried to our setting as well.
\end{remark}

\subsection{Relation to other results in the literature}
\label{section-review}

In general, works that study long-run dynamics of asset markets based on
ideas of natural selection of investment strategies can be attributed to the
field of \emph{Evolutionary Finance}, which has been developed since the
1990-2000s. Recent reviews (mostly of discrete-time models) can be found in
\cite{EvstigneevHens+16,Holtfort19}.

Let us first mention other works related to the results of
\Cref{theorem-survival,,theorem-close,,theorem-dominate}. A similar model in discrete time
was studied in the paper \cite{AmirEvstigneev+13}. Its main difference is
that at every moment of time the whole wealth is reinvested, i.e.\ the wealth
equation, instead of \eqref{wealth-discrete-1}, is the following one:
\begin{equation}
Y_{t}^m = \sum_n\frac{\lambda_t^{m,n} Y_{t-1}^m}{\sum_k
\lambda_t^{k,n}Y_{t-1}^k}A_t^n.
\label{AESH-capital}
\end{equation}
Notice that it can be formally obtained from \eqref{wealth-discrete-1} by
taking $\delta_t=1$. The main results of the paper \cite{AmirEvstigneev+13}
also consist in finding a survival strategy in an explicit form and proving
that all survival strategies are asymptotically close to it. In that model,
a survival strategy is defined by the same formula as
\eqref{dicrete-survival} with $\delta_t=1$ (one should put $v\equiv 1$
in~\eqref{a-def}). As extension of the model was considered in the paper
\cite{DrokinZhitlukhin19}, where investors can also decide what part of
their wealth they allocate for investment in the assets, and what part they
keep in a risk-free account. This is expressed in that
$\sum_n\lambda_t^{m,n}$ may be less than 1.

Similar results were also obtained in the paper \cite{AmirEvstigneev+11} for
a (more difficult) model in discrete time which assumes that investors can
sell their assets at subsequent moments of time for the price determined by
the market (through the balance of supply and demand) -- such an assumption
is natural for a model of a stock market. Quite remarkably, a survival
strategy in that model also exists and can be found in the class of basic
strategies which depend only on the structure of dividend sequences, but not
on actions of investors.

Let us also mention the paper \cite{BlumeEasley92} (see also the subsequent
paper \cite{Blumeeasley06}) -- one of the first in this direction -- where a
result similar to our \Cref{theorem-survival} was obtained (among other results).
The model considered in that paper is a simple particular case of
\eqref{AESH-capital}, where at each moment of time only one asset yields a
payoff and its amount, if paid, is known in advance.

Note that the model \eqref{AESH-capital} cannot be straightforwardly
generalized to the case of continuous time, since a continuous-time model
should allow that during an ``infinitesimally short'' period of time the
payoff $A_t$ can be ``infinitesimally small'' -- but then equation
\eqref{AESH-capital} makes no sense. There is no such problem in our model
due to the assumption $\delta_t<1$.

Among (few) other results for the case of continuous time, let us mention
the paper \cite{PalczewskiSchenkHoppe10-JEDC}, where convergence of a
discrete-time model to a continuous-time model was studied, and the paper
\cite{PalczewskiSchenkHoppe10-JME}, where questions of survival and
dominance of investment strategies were investigated in the case when
payoffs are specified by absolutely continuous non-decreasing processes.

Finally, with regard to \Cref{theorem-opt-growth}, one can see that survival
strategies, and in particular $\hat\lambda$, are similar to growth optimal
strategies (also called log-optimal strategies, benchmark strategies,
numéraire portfolios) in models of mathematical finance without competition,
in the sense that they lead to the fastest growth of wealth. An account on
growth optimal strategies can be found, for example, in
\cite[Chapter~5]{CoverThomas12}, \cite{HakanssonZiemba95} for discrete-time
models, and in \cite{KaratzasKardaras07,PlatenHeath06} for general
continuous-time models. However, note that there is an essential difference
between survival strategies and growth optimal strategies: the latter are
obtained as solutions of single-agent optimization problems for wealth
processes, while the former cannot be obtained in such a way because
competitors’ strategies are unknown to an investor.

\section{Proofs}
\label{section-proofs}

Before we proceed to the proofs, let us briefly recall, for the reader's
convenience, the notion of the stochastic exponent, which will be used in
several places below. If $Z$ is a scalar semimartingale, then the stochastic
exponent $\e(Z)$ of $Z$ is the semimartingale that solves the stochastic
differential equation (which always has a unique strong solution, see
\cite[\Ss\,I.4f]{JacodShiryaev02})
\[
d \e(Z)_t = \e(Z)_{t-} d Z_t, \qquad \e(Z)_0=1.
\]
In all the cases we are going to consider, only adapted \cadlag\ processes
with finite variation will be used as $Z$, so this equation should be
understood in the sense of pathwise Lebesgue--Stieltjes integration. The
Dolean--Dade formula implies that in this case
\begin{equation}
\e(Z)_t = e^{Z_t} \prod_{s\le t} (1+\Delta Z_s) e^{-\Delta Z_s}.
\label{dolean-dade}
\end{equation}
In particular, if $\Delta Z>-1$, then $\e(Z) > 0$ and $\e(Z)_- > 0$.

\bigskip
\noindent
\textbf{Proof of \Cref{proposition-correctness}.} Introduce the function
$F\colon \R_+^{MN+ M}\to\R^{MN}_+$ which specifies the distribution of
payoffs in equation \eqref{wealth}:
\[
[F(\lambda,y)]^{m,n} = \frac{\lambda^{m,n} y^m}{\sum_k \lambda^{k,n}
y^k}.
\]
When the denominator equals 0 for some $n$, we define $[F(\lambda,y)]^{m,n}
= 1/M$. It is straightforward to check that
$|\partial[F(\lambda,y)]^{m,n}/\partial y^k| \le 1/y^k$. Hence, $F$ is
Lipschitz continuous in $y$ on any set $\{y:y^m\ge a\ \text{for all}\ m\}$,
where $a$ is a positive constant. Let $l(a)$ be a function, defined for
$a\in(0,\infty)$, such that $|F(\lambda,y) - F(\lambda,\tilde y)|\le
l(a)|y-\tilde y|$ for any $\lambda\in\R^{MN}_+$ and $y,\tilde y\in \R_+^M$
with $y^m,\tilde y^m \ge a$ for all $m$. We can assume that $l(a)\ge 1$ for
any $a$ and $l(a)$ is bounded on any compact set in $\R_+\setminus\{0\}$.

Let $y^* = \min_m Y_0^m$. Define the sequence of stopping times $\tau_i$
with $\tau_0=0$ and, for $i\ge1$,
\[
\begin{split}
\tau_i = \inf\biggl\{&t\ge \tau_{i-1} : |X_t| \ge
|X_{\tau_{i-1}}|+\frac{1}{4l(y^* \e(-V)_{\tau_{i-1}}/2)}\quad \text{or}\\
&V_t \ge V_{\tau_{i-1}} + \frac14\wedge
\frac{y^* \e(-V)_{\tau_{i-1}}}{2(y^* + |X_{\tau_{i-1}}| + 1/4)}\biggr\}
\wedge (\tau_{i-1}+1),
\end{split}
\]
where $\inf\emptyset = \infty$. It is not hard to see that $\tau_i\le i$ and
$\tau_i\to\infty$ as $i\to\infty$.

We will construct a solution of \eqref{wealth} by induction on the
intervals $[0,\tau_i]$. Namely, we will define a sequence of adapted
\cadlag\ processes $Y^{(i)}$ with finite variation on any interval $[0,t]$
that satisfy equation \eqref{wealth} on $[0,\tau_i]$ and have the property
\[
Y^{(i)}_t = Y^{(i-1)}_t\ \text{for}\ t\le \tau_{i-1}.
\]

For $i=0$, let $Y^{(0)}_t = Y_0$ for all $t\ge0$. Suppose the process
$Y^{(i-1)}$ is constructed. Observe that equation~\eqref{wealth} implies
that for each $m$
\begin{equation}
Y_0^m \e(-V)_t \le Y_t^{(i-1),m} \le Y_0^m + |X_t|\ \text{for}\ t\le
\tau_{i-1}.\label{proof-Y-bound}
\end{equation}
Here, the right inequality is clear, and the left one follows from that the
process $Y^{(i-1),m}/\e(-V)$ is non-decreasing, which can be seen by
computing its stochastic differential.

Let us now construct $Y^{(i)}$. Consider the Banach space of bounded
\cadlag\ functions $f\colon \R_+\to\R^M$ with the norm $\|f\| = \sup_{t\ge
0} |f_t|$ and denote by $\mathbb{D}$ is closed subset consisting of $f$ with
values in $\R_+^M$. For each $\omega$, consider the operator $H$ which maps
$f\in \mathbb{D}$ to
\begin{equation}
\begin{aligned}
H(\omega,f)_t &= Y^{(i-1)}_{t\wedge\tau_{i-1}}  \\&+
\I(t> \tau_{i-1})\int_{(\tau_{i-1},t]}\I(u< \tau_i) 
(F(\lambda_u, f_{u-}) d X_u - f_{u-} d V_u) ,
\end{aligned}
\label{proof-H}
\end{equation}
where the random variables in the right-hand side are evaluated for a
given~$\omega$. Notice that $H$ preserves the adaptedness of processes in
the sense that if $Y$ is an adapted process, then so is $H(Y)$.

For each $\omega$, introduce the set
\[
\begin{split}
\mathbb{D}^{(i)}(\omega) = \Bigl\{&f\in \mathbb{D} : \frac12 Y_0^m
\e(-V)_{\tau_{i-1}}(\omega) \le f^m_t\le Y_0^m +
|X_{\tau_{i-1}}(\omega)|+\frac14\\ &\text{for all}\ m\ \text{and}\
t\ge \tau_{i-1}(\omega)\Bigr\}.
\end{split}
\]
Let us show that $H(\omega)$ maps $\mathbb{D}^{(i)}(\omega)$ into itself.
Indeed, a function $H(f)$ is \cadlag. The upper bound for $H(f)$ in the
definition of $\mathbb{D}^{(i)}$ follows from the inequalities (for $t\ge
\tau_{i-1}$)
\[
H(f)_t^m \le Y_{\tau_{i-1}}^m + |X_{\tau_i-}| - |X_{\tau_{i-1}}| \le Y_0^m +
|X_{\tau_i-}| + \frac14,\label{proof-1-upper-bound}
\]
where the first inequality here follows from~\eqref{proof-H} with the bound
$|F(\lambda,y)|\le 1$, and the second one follows from the right inequality
in~\eqref{proof-Y-bound} and the estimate $|X_{\tau_i-}| - |X_{\tau_{i-1}}|
\le 1/4$ which holds by the choice of $\tau_i$.

The lower bound for $H(f)$ follows from that for $t\ge \tau_{i-1}$
\[
\begin{aligned}
H(f)^m_t &\ge Y^m_{\tau_{i-1}}- \int_{(\tau_{i-1},\tau_i)} f_{u-} d V_u \\ &\ge
Y_0^m \e(-V)_{\tau_{i-1}} - \Bigl(Y_0^m + |X_{\tau_{i-1}}(\omega)|+\frac14\Bigr)
(V_{\tau_i-} - V_{\tau_{i-1}}) \\&\ge \frac12 Y_0^m \e(-V)_{\tau_{i-1}},
\end{aligned}
\]
where the second inequality holds due to \eqref{proof-Y-bound} and the upper
bound for $f\in \mathbb{D}^{(i)}$, while the third inequality is valid
according to the choice of $\tau_i$.

Thus $H$ maps $\mathbb{D}^{(i)}$ into itself. Moreover, it is a contraction
mapping since for any $f,\tilde f \in \mathbb{D}^{(i)}$
\[
\begin{split}
\|H(f) - H(\tilde f)\| &\le \int_{(\tau_{i-1},\tau_i)}(
|F(\lambda_t, f_{t-}) - F(\lambda_t, \tilde f_{t-})|d X_t +|f_{t-} - \tilde
f_{t-}|d V_t )  \\&\le
\frac12\|f-\tilde f\|.
\end{split}
\]
Here, in order to bound the integral with respect to $dX_t$, we use that
$f_{t-},\tilde f_{t-} \ge y^*\e(-V)_{\tau_{i-1}}/2$ by the definition of
$\mathbb{D}^{(i)}$, hence the integrand can be bounded by $|f_{t-}-\tilde
f_{t-}|l(y^*\e(-V)_{\tau_{i-1}}/2)$, and so the value of the integral does
not exceed $\|f-\tilde f\|/4$ because $|X_{\tau_i-}| - |X_{\tau_{i-1}}| \le
1/(4l(y^*\e(-V)_{\tau_{i-1}}/2))$. In a similar way, the integral with
respect to $d V_t$ is also not greater than $\|f-\tilde f\|/4$ because
$V_{\tau_i-} - V_{\tau_{i-1}} \le 1/4$.

Consequently, $H$ has a fixed point $\tilde Y$, which satisfies equation
\eqref{wealth} on the half-interval $[0,\tau_i)$. The process $\tilde Y$ is
adapted, since it can be obtained as the limit (for each $\omega$ and $t$)
of the adapted processes $H^n(Y^{(i-1)})$ as $n\to\infty$, where $n$ stands
for $n$-times application of $H$. Define
\[
Y^{(i)}_t = \tilde Y_{t} \I(t<\tau_i) +
\left[\tilde Y_{\tau_i-} +  F(\lambda_{\tau_i}, \tilde Y_{\tau_i-}) \Delta
X_{\tau_i} - \tilde Y_{\tau_i-}\Delta V_t\right]
\I(t\ge \tau_i).
\]
Then $Y^{(i)}$ is the sought-for process which satisfies \eqref{wealth} on
the whole interval $[0,\tau_i]$. The strict positivity of $Y^{(i)}$ and
$Y^{(i)}_-$ follows from the left inequality in \eqref{proof-Y-bound}.

The uniqueness of the solution of \eqref{wealth} follows from the
uniqueness of the fixed point of the operator $H$ on each step of the
induction.
\qed

\bigskip
\noindent
\textbf{Proof of \Cref{prop-equal-realizations}.} Suppose $G'$ is a
predictable process satisfying \eqref{operation-time} with some process $b'$
and transition kernel $K'$. Then the random measure generated by $G$ on
$\R_+$ is a.s.\ absolutely continuous with respect to the measure generated
by $G'$, and, according to \cite[Proposition~I.3.13]{JacodShiryaev02},
there exists a non-negative predictable process $\rho$ such that $G = \rho
\cint G'$. Hence $X^c=b\cint G = (\rho b)\cint G'$, while, at the same time,
$X^c = b'\cint G'$, so we have
$\rho b=b'$ ($\P\otimes G'$-a.s.\ and, hence, $\Q$-a.s.). In a
similar way,
$\rho K = K'$ ($\Q$-a.s.), which implies $\rho a = a'$ ($\Q$-a.s.), where
$a,a'$ are the processes defined by~\eqref{a-def} with respect to $K$ and
$K'$. Then, by \eqref{survival}, we have $\hat\lambda = \hat\lambda'$
($\Q$-a.s.), which proves the first claim of the
\namecref{prop-equal-realizations}.

To prove the second claim, observe that if $f_t, f_t'$ are predictable
non-negative processes such that $f=f'$ ($\Q$-a.s.), then $f\cint X = f'\cint
X$. Hence, if in the wealth equation
\eqref{wealth-sde} we replace the strategy $\lambda^1 = \hat\lambda$ by
$\hat\lambda'$, the wealth process $Y_t(Y_0,\Lambda)$ will 
remain its solution (up to $\P$-indistinguishability).
\qed

\bigskip
\noindent
\textbf{Proof of \Cref{H-bound}.}
Define the sequence of stopping times $\tau_i$, $i\ge 0$, with $\tau_0=0$
and
\[
\tau_i = \inf\{t\ge \tau_{i-1} : W_{t}\le 1/i\ \text{or}\ W_t\ge i\}
\wedge (\tau_{i-1}+1),\qquad i\ge1.
\]
Then we have
\[
H_{\tau_i-} \le i(|a|+|b|)\cint G_{\tau_i-} \le
i\biggl(i\wedge \frac{|x|}{1-\gamma_V}\biggr)*\nu_{\tau_i-} + i|X^c_{\tau_i-}| < \infty,
\]
where in the second inequality we used the bound $|a_t| \le \int_{\R_+^{N+1}
} (i \wedge |x|/(1-\gamma_V)) K_t(dx,dv)$ for $t< \tau_i$, which follows
from~\eqref{a-def}, and in the last inequality used the property $(|x|\wedge
1)* \nu_t < \infty$. Since $W>0$ and $W_->0$, we have $\tau_i\to\infty$ as
$i\to\infty$, and, hence, $H_t<\infty$ for all $t\ge 0$.
\qed

\bigskip In the following \namecref{lemma-logsum} we establish an auxiliary
inequality that will be used in the subsequent proofs. To state it, let us
introduce the function
\[
\lln x = 
\begin{cases}
\ln x,&\text{if}\ x>0,\\
-1, &\text{if}\ x\le 0.
\end{cases}
\]

\begin{lemma}
\label{lemma-logsum}
Suppose vectors $\alpha,\beta\in \R^N_+$ are such that
$|\alpha|=|\beta|=1$, and, for each $n$, we have $\alpha^n=0$ if  $\beta^n=0$. Then
\begin{equation}
\alpha(\lln\alpha - \lln \beta) \ge \frac{\|\alpha-\beta\|^2}{4} .\label{inf-ineq}
\end{equation}
\end{lemma}

\noindent
\textbf{Proof.} For vectors with strictly positive coordinates this
inequality follows from the inequality for the Kullback--Leibler and
Hellinger--Kakutani distances (a direct proof can be found, for example, in
\cite{AmirEvstigneev+13}, Lemma 2): it is sufficient to consider
$\alpha,\beta$ as probability distributions on a set of $N$ elements. For
vectors which may have null coordinates, instead of $\alpha,\beta$ one can
take $c\alpha+(1-c)u$, $c\beta+(1-c)u$, where $c\in(0,1)$ and $u$ is a
vector with strictly positive coordinates and $|u|=1$. Then let $c\to1$ and
use the continuity of the function $x\lln x$ and the norm $\|\,\cdot\,\|$ to
obtain \eqref{inf-ineq}.
\qed

\bigskip
\noindent
\textbf{Proof of \Cref{theorem-survival}.}
Suppose investor $m$ uses a strategy $\lambda$ which satisfies conditions
(a)--(c). First we are going to prove that the process
$S_t = \ln r_t^m + U\cint H_t$ is a local submartingale.

It will be convenient to represent the wealth process $Y^m$ and the process
of total wealth $W$ as stochastic exponents. Put, for brevity,
\[
\theta_t = \frac{1}{W_{t-}},
\]
and associate with the strategy $\lambda^m=\lambda$ of  investor $m$ the
$N$-dimensional predictable process $F_t$ with the components
\[
F_t^n = \frac{\lambda_t^{m,n}\theta_t }{\sum_k \lambda_t^{k,n} r_{t-}^k},
\]
where in the case of the indeterminacy $0/0$ we define $F_t^n=(MW_{t-}r_{t-}^m)^{-1}$.
Then the processes $Y_t^m$ and $W_t$ satisfy the equations
\[
d Y_t^m = Y_{t-}^m (F_t d X_t - dV_t), \qquad
d W_t = d | X_t|  -W_{t-} dV_t.
\]
Therefore, they can be represented in the form
\[
Y^m_t = Y_0^m\e(F\cint X - V)_t,\qquad
W_t = W_0\e(\theta \cint|X| - V)_t.
\]

Let $Z_t = \ln r^m_t = \ln \e(F\cint X-V)_t - \ln
\e(\theta\cint|X|-V)_t + Z_0$. As follows from \eqref{dolean-dade},
\[
Z_t =
(F-\theta)\cint X_t^c + \sum_{s\le t} \ln\left( \frac{1 -\Delta V_s +F_s\Delta
X_s}{1 -\Delta V_s +\theta_s|\Delta X_s|} \right) +Z_0
\]
(in the first term, $\theta$ is subtracted from each coordinate of $F$). Define
the predictable function $f\colon \Omega\times\R_+\times\R^{N+1}_+ \to \R$ by
\[
f(\omega,t,x,v) = \ln\left( \frac{1 -v+ F_t(\omega)x }{1-v+\theta_t(\omega)|x| } \right).
\]
Using this function, it is possible to write
\[
Z_t = (F-\theta)\cint X_t^c + f*\mu_t +Z_0.
\]
Let us prove the representation 
\begin{equation}
Z_t = g \cint G_t + f * (\mu-\nu)_t + Z_0\label{submart}
\end{equation}
with the function
\[
g_t = (F_t-\theta_t) b_t + \int_{\R^{N+1}_+} f_t(x,v) K_t(dx,dv).
\]
To prove \eqref{submart}, it is sufficient to show that $f*\nu_t
<+\infty$ and $g\cint G_t> -\infty$ for all $t$ (and then we will also have
$|f|*\nu_t < \infty$).

Consider the stopping times $\tau_i = \inf\{t\ge 0 : r^m_t \le 1/i\
\text{or}\ W_t\le 1/i\}$ with $\inf\emptyset = +\infty$. It is not difficult
to see that $F_t^n \le \theta_t(r^m_{t-})^{-1}$ for all $n$.
Then $f \le \frac{i^2|x|}{1-\gamma_V+i|x|}$ on the set $\{t<
\tau_i(\omega)\}$. Since $(|x|\wedge 1)*\nu_t <\infty$ for all $t$, we
have $f*\nu_{\tau_i-} <+\infty$. Because $\tau_i\to \infty$ as
$i\to\infty$ (due to the strict positivity of $W$, $W_-$, $r^m$, and
$r_-^m$), passing to the limit $i\to\infty$, we obtain $f*\nu_t <+\infty$
for all $t$.

Let us prove that $g\cint G_t> -\infty$ for all $t$. Define the set
$\mathcal{X}(\omega,t) = \{(x,v)\in \R^{N+1}_+\setminus\{0\} : x^n = 0
\text{ if }F_t^n(\omega)=0,\; n=1,\ldots,N\}$. Using Jensen's inequality and
the concavity of the logarithm, we find that for any
$(x,v)\in\mathcal{X}(\omega,t)$
\begin{multline*}
f_t(x,v) = \ln \left( \frac{1-v}{1-v+\theta_t|x|} + \frac{\theta_t|x|}{1-v+\theta_t|x|}
\frac{F_tx}{\theta_t|x|}\right) \\\ge \frac{\theta_t|x|}{1-v+\theta_t|x|}
\ln\left(\frac{F_tx}{\theta_t|x|}\right) \ge \frac{\theta_tx\lln (F_t/\theta_t)}{1-v+\theta_t|x|}.
\end{multline*}
This implies that for each $t$
\[
\int_{\R^{N+1}_+} f_t(x,v)K_t(dx,dv) = \int_{\mathcal{X}_t} f_t(x,v)
K_t(dx,dv) \ge \theta_t a_t\lln (F_t/\theta_t),
\]
where we use that $K_t(\R^{N+1}_+ \setminus \mathcal{X}_t) = 0$. Indeed, the
set $\R^{N+1}_+ \setminus \mathcal{X}(\omega,t)$ consists of $(x,v)$ such
that $F_t^n(\omega)=0$ but $x^n>0$ for some $n$. On the set $\{F_t^n = 0\}$
we have $\lambda_t^{m,n}=0$, so, by condition (a) of the theorem,
$\hat\lambda_t^n = 0$ a.s.\ on this set, and therefore
$K_t(\{x^n>0\}) = 0$.

Then we can write
\begin{equation}
\begin{split}
g_t &\ge (F_t-\theta_t)b_t + \theta_t a_t \lln (F_t/\theta_t) \ge \theta_t
(a_t+b_t) \lln (F_t/\theta_t) \\ &\ge \hat\lambda_t (\lln \lambda_t - \lln
\pi_t) |a_t+b_t|\theta_t
\end{split}
\label{g-ineq}
\end{equation}
(in the second inequality we used that $F_t^n/\theta_t-1 \ge \lln
(F_t^n/\theta_t)$), where
\begin{equation}
\pi_t^n = \sum_k \lambda_{t}^{k,n} r_{t-}^k.\label{pi-t}
\end{equation}
Note that $|\pi_t| =1$. Applying \Cref{lemma-logsum} to the vectors
$\alpha=\hat\lambda_t$ and $\beta=\pi_t$, from formula \eqref{g-ineq} we
find
\[
g_t \ge \hat\lambda_t(\lln \lambda_t - \lln\hat\lambda_t)|a_t+b_t|\theta_t =  - U_t|a_t+b_t|\theta_t,
\]
where, to obtain the equality, we changed $\lln$ to $\ln$, which is possible
due to condition~(a). Then, by condition (b), $g\cint G_t \ge -U\cint H_t>
-\infty$, which proves representation \eqref{submart}. In particular,
$|f|*\nu_t<\infty$ for any $t$. Since a predictable non-decreasing
finite-valued process is locally integrable
\cite[Lemma~1.6.1]{LiptserShiryaev89en}, the process $|f|*\nu_t$ is locally
integrable, and, hence, $f*(\mu-\nu)_t$ is a local martingale.
Therefore, $S_t=Z_t+ U\cint H_t$ is a local submartingale. Following a
standard technique, let us show that this fact and condition (c)
imply that $Z_t$ has an a.s.-finite limit as $t\to\infty$.

Consider the sequence of stopping times
\begin{equation}
\tau_i = \inf\{t\ge 0: U\cint H_t \ge i\}, \qquad
i\in\mathbb{N},\label{tau-i}
\end{equation}
where $\inf\emptyset = +\infty$. By condition (b), for a.a.\ $\omega$ we
have $\tau_i(\omega)=\infty$ starting from some $i$. For each $i$, the
process $S^{(i)}_t = S_{t\wedge \tau_i}$, $t\ge 0$, is a local
submartingale and, moreover, for all $t\ge 0$
\begin{equation}
S^{(i)}_t \le U\cint H_{\tau_i} \le i + U_{\tau_i}\Delta H_{\tau_i}\I(\tau_i
< \infty).\label{S-bound}
\end{equation}
From condition (c), it follows that the random variable in the right-hand
side of the above inequality is integrable. Consequently, $S^{(i)}_t$ is a
usual submartingale and there exists the a.s.-finite limit
$\lim_{t\to\infty} S^{(i)}_t = S_{\tau_i}$ (by Doob's martingale convergence
theorem, see Theorem I.1.39 in \cite{JacodShiryaev02}). Letting
$i\to\infty$, we obtain the existence of the a.s.-finite limit
$S_\infty=\lim_{t\to\infty} S_t$ and $Z_\infty = S_\infty - U\cint
H_\infty$. This implies $\lim_{t\to\infty} r_t^m = \exp(Z_\infty) > 0$,
which proves the theorem.

\bigskip
\noindent
\textbf{Proof of \Cref{prop-survival-sufficient}.} It is clear that if the conditions
of the \namecref{prop-survival-sufficient} are satisfied then the strategy $\lambda$
satisfies condition (a). Denote $\hat\xi = \inf_{t,n} \hat \lambda_t^n$ and
$\xi = \inf_{t,n} \lambda_t^n$. Then we have the inequalities
\begin{align*}
&(\lambda_t^n - \hat \lambda_t^n)(\ln \lambda_t^n - \ln\hat\lambda_t^n) \le
\frac{(\lambda_t^n-\hat \lambda_t^n)^2}{\hat \lambda_t^n} \quad
\text{ if }\lambda_t^n\ge \hat\lambda_t^n,\\
&(\lambda_t^n - \hat \lambda_t^n)(\ln \lambda_t^n - \ln\hat\lambda_t^n) \le
\frac{\ln(\xi)(\lambda_t^n-\hat \lambda_t^n)^2}{(\xi-1)\hat \lambda_t^n} \quad
\text{ if }\lambda_t^n< \hat\lambda_t^n,
\end{align*}
where we used the inequalities $\ln(1+x)\le x$ if $x\ge 0$ and
$\ln(1+x)\ge x\epsilon^{-1}\ln(1+\epsilon)$ if $x\in[\epsilon,0]$, $\epsilon>-1$,
applied to $x= (\lambda_t^n-\hat\lambda_t^n)/\hat\lambda_t^n$ and
$\epsilon=\xi-1$. Since $\lambda_t(\ln \lambda_t - \ln\hat\lambda_t)\ge 0$
(by \Cref{lemma-logsum}), we find
\[
U_t \le \frac{\ln(\xi)\|\lambda_t-\hat\lambda_t\|^2}{(\xi-1)\hat\xi}.
\]
Consequently, $U\cint H_\infty < \infty$, so $\lambda$ satisfies
condition (b). \qed

\bigskip
In order to prove \Cref{theorem-close,theorem-dominate}, we
will need the following auxiliary result.  

\begin{lemma}
\label{lemma-diff-bound}
Suppose investor $m$ uses a strategy $\lambda$ satisfying conditions
(a)--(c) of \Cref{theorem-survival}, and let $\pi_t$ be the process defined
by \eqref{pi-t}. Then $\|\hat\lambda - \pi\|^2 \cint H_\infty < \infty$.
\end{lemma}

\noindent
\textbf{Proof.} In the course of proof of \Cref{theorem-survival} we have
established inequality~\eqref{g-ineq}. Together with \eqref{inf-ineq}, it
implies
\[
\|\hat\lambda-\pi\|^2\cint H_\infty \le
4\hat\lambda(\lln\hat\lambda-\lln\pi) \cint H_\infty \le 4(g\cint G_\infty
+U\cint H_\infty).
\]
It remains to show that $g\cint G_\infty + U\cint H_\infty < \infty$.
Consider the stopping times~$\tau_i$ defined in \eqref{tau-i}. Then for any
$t\ge0$
\[
\E (g\cint G_{t \wedge\tau_i} +U\cint H_{t \wedge\tau_i}) = \E S_{t}^{(i)}
+ S_0
\le i + \E(U_{\tau_i} \Delta H_{\tau_i}\I(\tau_i<\infty)),
\]
where the equality holds because $g\cint G_{t\wedge \tau_i}+U\cint
H_{t\wedge \tau_i}$ is the compensator of the submartingale $S^{(i)}$
defined in the proof of \Cref{theorem-survival}, and the inequality holds in view
of~\eqref{S-bound} and that $S_0=Z_0\le 0$. Passing to the limit
$t\to\infty$, by the monotone convergence theorem $\E (g\cint G_{\tau_i}
+U\cint H_{\tau_i})<\infty$, and hence $g\cint G_{\tau_i} +U\cint
H_{\tau_i}<\infty$. Passing to the limit $i\to\infty$, we obtain $g\cint
G_{\infty} +U\cint H_{\infty}<\infty$, since $\tau_i(\omega)=\infty$
starting from some $i$ for a.a.\ $\omega$. \qed

\bigskip
\noindent
\textbf{Proof of \Cref{theorem-close}.} Consider the market where investor 1
uses the strategy $\lambda^1=\lambda$, and the other investors use the
strategy $\hat\lambda$, i.e.\ $\lambda^m = \hat\lambda$, $m=2,\ldots,M$. In
this case $\pi_t = \lambda_t r^1_{t-} + \hat\lambda_t (1-r^1_{t-})$ and
$\|\hat\lambda - \pi_t\| = r_{t-}^1\|\hat\lambda_t-\lambda_t\|$. Then from
\Cref{lemma-diff-bound} we obtain $(r_{-}^1 \|\hat\lambda - \lambda\|)^2
\cint H_\infty < \infty$. Since $\lambda$ is a survival strategy, we
have $\inf_t r_t^1 > 0$. Therefore, $\|\hat\lambda - \lambda\|^2\cint
H_\infty <\infty$. \qed

\bigskip
\noindent
\textbf{Proof of \Cref{theorem-dominate}.} We have $\pi_t = \lambda_t r_{t-}^m+
(1-r_{t-}^m) \tilde\lambda_t$. By virtue of~\Cref{lemma-diff-bound}, we
obtain
\[\|\hat \lambda - \lambda + (1-r^m_-)(\lambda-\tilde
\lambda)\|^2 \cint H_\infty = \|\hat \lambda - \pi\|^2\cint H_\infty <
\infty.
\]
Since the strategy $\lambda$ is survival, $\|\hat\lambda-\lambda\|^2\cint
H_\infty <\infty$ by \Cref{theorem-close}. From these inequalities, it
follows that $(1-r^m_-)^2\|\lambda-\tilde \lambda\|^2 \cint H_\infty <
\infty$. According to \Cref{theorem-survival}, there exists the a.s.-finite limit
$r^m_\infty=\lim_{t\to\infty} r_t^m$. Then necessarily $r_\infty^m =1$
a.s.\ on the set $\{\|\lambda-\tilde \lambda\|^2\cint H_\infty =\infty\}$,
which a.s.\ coincides with the set $\{\|\hat\lambda-\tilde \lambda\|^2\cint
H_\infty =\infty\}$ as follows from \Cref{theorem-close}. \qed

\bigskip
\noindent
\textbf{Proof of \Cref{theorem-opt-growth}.} 1) If investor $m$ uses a
survival strategy, then for any strategies of the other investors the
inequality $\inf_t r_t^m>0$ holds with probability one. Then $\sup_t
W_t/Y_t^m < \infty$ and therefore $\sup_t Y_t^k/Y_t^m < \infty$ for any $k$.
Hence we obtain the inequality
\[
\limsup_{t\to \infty} \frac{1}{t}\ln \frac{Y_t^k}{Y_t^m} \le 0,
\]
which easily implies the first claim of the \namecref{theorem-opt-growth}.

2) From the proof of \Cref{theorem-survival}, it follows that if investor $m$ uses
the strategy $\hat\lambda$, then $\ln r_t^m$ is a submartingale, so for any
$s\le t$
\begin{equation}
\E (\ln r_t^m \mid \F_s) \ge \ln r_s^m.\label{proof-4.1}
\end{equation}
Using that $\E(\ln W_t \mid \F_s)$ is finite-valued due to the assumption of
the \namecref{theorem-opt-growth}, and adding to the both sides of above the
inequality $\E(\ln(W_t/Y_s^m) \mid \F_s)$ we obtain
\begin{equation}
\E \biggl(\ln\frac{Y_t^m}{Y_s^m} \midd \F_s\biggr) \ge
\E\biggl(\ln\frac{W_t}{W_s} \midd \F_s
\biggr).\label{proof-4.2}
\end{equation}
Let $\tilde r_t = \tilde Y_t/W_t = 1-r_t^m$. From \eqref{proof-4.1}, by
Jensen's inequality, we find
\[
\E (\ln \tilde r_t \mid \F_s) \le \ln \tilde r_s,
\]
where the conditional expectation may assume the value $-\infty$. Then,
similarly to \eqref{proof-4.2}, we have
\[
\E \biggl(\ln\frac{\tilde Y_t}{\tilde Y_s} \midd \F_s\biggr) \le
\E\biggl(\ln\frac{W_t}{W_s} \midd \F_s
\biggr),
\]
which together with \eqref{proof-4.2} proves the second claim of the
\namecref{theorem-opt-growth}.
\qed

\bibliographystyle{abbrv}
\bibliography{survival}
\end{document}